\documentstyle[12pt,epsf]{article}

\def\gtrsim{\raisebox{-4pt}{$\stackrel{\textstyle >}{\sim}$}}

\begin{document}
\hspace\parindent
\thispagestyle{empty}

\bigskip
\bigskip
\bigskip

\centerline{\LARGE \bf Numerical Analysis of Dynamical Systems}
\centerline{\LARGE \bf and the Fractal Dimension of Boundaries}

\bigskip
\bigskip
\bigskip

\begin{center}
{\large
L. G. S. Duarte,
L. A. C. P. da Mota,
H. P. de Oliveira,
}

{\large R. O. Ramos
and J. E. F. Skea}
\footnote{E-mails: lduarte@dft.if.uerj.br, damota@dft.if.uerj.br,
                   oliveira@dft.if.uerj.br, rudnei@dft.if.uerj.br and
                   jimsk@dft.if.uerj.br}
\end{center}

\bigskip

\centerline{\it Universidade do Estado do Rio de Janeiro,}
\centerline{\it Instituto de F\'{\i}sica, Depto. de F\'{\i}sica Te\'orica,}
\centerline{\it 20550-013 Rio de Janeiro, RJ, Brazil}

\bigskip
\bigskip

\begin{abstract} 
A set of MapleV R.4/5 software routines for calculating the numerical
evolution of dynamical systems and flexibly plotting the results 
is presented.  The package consists of an initial condition
generator (on which the user can impose quite general constraints), a
numerical solving manager, plotting commands that allow the user to
locate and focus in on regions of possible interest and, finally, a set
of routines that calculate the fractal dimension of the boundaries of
those regions. A special feature of the software routines presented here
is an optional interface in C, permitting fast numerical integration
using standard Runge-Kutta methods, or variations, for high precision
numerical integration.
\end{abstract}

\bigskip
\centerline{ \underline{\hspace{6.5 cm}} }

\medskip

\centerline{ {\bf (Submitted to Computer Physics Communications)} }

\bigskip
\bigskip
\bigskip
\bigskip
\bigskip
\bigskip


\newpage
\bigskip
\hspace{1pc}
{\bf PROGRAM SUMMARY}
\bigskip

\begin{footnotesize}
\noindent
{\em Title of the software package:} Ndynamics.   \\[10pt]
{\em Catalogue number:} (supplied by Elsevier)                \\[10pt]
{\em Software obtainable from:} CPC Program Library, Queen's
University of Belfast, N. Ireland (see application form in this issue)
\\[10pt]
{\em Licensing provisions:} none  \\[10pt]
{\em Operating systems under which the program has been tested:}
Linux (RedHat 5.2 and Debian 2.0.34), Windows 95, Windows 98.
\\[10pt]
{\em Programming languages used:} Maple V Release 4 and 5 and ANSI C.
\\[10pt]
{\em Memory required to execute with typical data:}  32 Megabytes. \\[10pt]
{\em No. of lines in distributed program, including On-Line Help,
etc.:} 1370.                                                   \\[10pt]
{\em Nature of mathematical problem}\\
Computation and plotting of numerical solutions of dynamical systems
and the determination of the fractal dimension of the boundaries. 
   \\[10pt]
{\em Methods of solution}\\
The default method of integration is a 5th order Runge-Kutta scheme,
but any method of integration present on the MAPLE system is available
via an argument when calling the routine. A box counting method is
used to calculate the fractal dimension of the boundaries.
   \\[10pt]
{\em Restrictions concerning the complexity of the problem}\\
Besides the inherent restrictions of numerical integration methods, this
first version of the package only deals with systems of first order
differential equations.
   \\[10pt]
{\em Typical running time}\\
This depends strongly on the dynamical system. With a Pentium II 450 PC
with 128 Mb of RAM, the integration of one graph (among the thousands
it is necessary to calculate to determine the fractal dimension)
takes from a fraction of a second to several seconds. The time for
plotting the graphs depends on the number of trajectories plotted.
If there are a few thousand, this may take 20 to 30 seconds.
   \\[10pt]
{\em Unusual features of the program}\\
This package provides user-friendly software tools for analyzing the
character of a dynamical system, whether it displays chaotic behavior,
etc.  Options within the package allow the user to specify characteristics
that separate the trajectories into families of curves. In conjunction
with the facilities for altering the user's viewpoint, this provides a
graphical interface for the speedy and easy identification of regions
with interesting dynamics.
An unusual characteristic of the package is its
interface for performing the numerical integrations in C using a
5th order Runge-Kutta method. This potentially improves the speed of the
numerical integration by some orders of magnitude and, in cases where it
is necessary to calculate thousands of graphs in regions of difficult
integration, this feature is very desirable.
\end{footnotesize}
\newpage
\hspace{1pc}
{\bf LONG WRITE-UP}

\section{Introduction}
\label{intro}


A large number of problems in physics, chemistry, etc, can be
represented in terms of a set of differential equations, which
constitute a (very often nonlinear) dynamical system.
Among these systems of interest, autonomous dynamical systems
are particularly common. A $d$-dimensional autonomous dynamical system can be
represented by

\begin{equation}
\frac{d\,{\bf X}(t)}{d\,t}={\bf F}({\bf X}),  
\label{syst}
\end{equation}   

\noindent 
where ${\bf X}=\left(X_1(t),X_2(t),\ldots,X_d (t) \right)$ and
${\bf F}=\left(F_1({\bf X}),F_2({\bf X}),\ldots,F_d ({\bf X}) \right)$. 
The $X_i(t)$, with
$i = 1,\ldots,d$, represent field variables, usually related to
quantities associated with the specific problem under
consideration, $t$ is a continuous parameter, which can often be
considered as the time, while the
functions $F_i({\bf X)}$, $i = 1,\ldots,d$, are general functions
of their arguments.
The $d$-dimensional space constituted by the $X_i(t)$ is called the system's
phase space, with trajectories in this phase space representing the
solutions of~(\ref{syst}). It is possible to interpret the
trajectories generated by all the initial conditions in phase space as
analogous to the paths followed by the particles of a flowing fluid with
velocity field $\dot{\bf X}$, where an overdot represents a derivative
with respect to $t$. If $F_i = F_i({\bf X},t)$, the system is said
to be a non-autonomous dynamical system. However, introducing a new
variable $X_{d+1} = t$, the $d$-dimensional non-autonomous dynamical
system can be transformed into a $(d+1)$-dimensional autonomous system.
An autonomous system is said to be Hamiltonian if (\ref{syst})
represents the set of Hamilton's equations and the phase space has
a simplectic structure. Otherwise we have a dissipative or non-Hamiltonian
system. In this case, volumes in phase space are not conserved, or,
equivalently,
${\bf \nabla}\cdot{\bf \dot{X}} = {\bf \nabla}\cdot{\bf F} \neq 0$.

Consider a general nonlinear autonomous dynamical system with dimension
$d \geq 3$. Among these systems those exhibiting chaotic behavior have
assumed 
particular interest and constitute a rule than an exception. Roughly
speaking, chaos means extreme sensitivity to small changes in the
initial conditions. Due to nonlinearity, fluctuations in the initial
conditions of chaotic systems evolve
such that they can completely alter the asymptotic outcome of the
unperturbed trajectories in phase space. There are several ways of
analyzing chaos in dynamical systems, such as the Liapunov
exponents or the technique of Poincar\'e surface of sections for
Hamiltonian systems. (For good reviews and books on chaotic
dynamical systems see, for instance \cite{caos1,caos2}.) We are
particularly interested in certain features underlying chaotic dynamical
systems, namely sets characterized by a non-integer dimension. Such
sets are known as fractals. In the realm of dissipative systems,
attractors with fractal properties have been called strange attractors,
whereas in Hamiltonian systems there exist the strange repellers \cite{repel},
identified as repellers with fractal properties. An important
numerical task in investigating chaotic behavior in dynamical systems is
therefore the determination of the fractal dimension~\cite{caos2}.

The presence of fractals is related to chaos, and so these
structures are also associated with sensitivity to small changes in initial
conditions. To show this relation, we follow closely the procedure
of~\cite{ott}. Let us consider a non-fractal basin boundary $\Sigma$ that
separates two attractors $A_-$ and $A_+$, usually fixed points (this
will be explained further in section~\ref{examples}). An uncertainty in
our initial
condition ${\bf X}(t_0) = {\bf X_0}$ that has strength $\epsilon$ means
that the initial condition lies somewhere in the region $|{\bf X} - {\bf
X_0}| \leq \epsilon$. According to Figure~1, given an uncertainty
$\epsilon$, the initial condition~1 always goes to the attractor $A_+$.
On the other hand, since the point~2 lies near the boundary $\Sigma$,
due to the uncertainty $\epsilon$ the actual trajectory may go to
either attractor $A_+$ or attractor $A_-$. The initial condition~1 is
said to be $\epsilon$-certain, and the initial condition~2 is
$\epsilon$-uncertain. Obviously, $\epsilon$-uncertain initial conditions
are those which lie within a distance $\epsilon$ of the basin boundary
$\Sigma$. Let us call $f(\epsilon)$ the probability of obtaining an
$\epsilon$-uncertain initial condition, in the sense that it is the
fraction of the $d$-dimensional volume of the phase space which lies
within $\epsilon$ of the boundary $\Sigma$. In the case of a simple
non-fractal boundary, $f(\epsilon) \sim \epsilon$.
This scaling law tells us that an improvement in the accuracy of
the initial condition
by, say, a factor $10$ (a reduction of $\epsilon$ by $10$),
reduces $f(\epsilon)$, and hence our probability of potential error, by a
factor of $10$. For the case of a fractal boundary the
scaling law~(see Appendix)

\begin{equation}
f(\epsilon) \sim \epsilon^\alpha \;
\label{scaling}
\end{equation}

\noindent 
is valid,
where $\alpha$ is the uncertainty exponent. The box-counting
dimension of the basin boundary, $D_0$, is given by~\cite{ott}

\begin{equation}
D_0 = d -\alpha \;,
\label{D_0}
\end{equation}

\noindent 
with $d$ denoting the dimension of the phase space. For a
fractal boundary, $D_0 > d -1$, implying that $\alpha < 1$, whereas for a
non-fractal boundary, $D_0 = d - 1$, and $\alpha = 1$. Suppose that, for
instance, $\alpha = 0.2$ and $f(\epsilon) \sim \epsilon^{0.2}$. Then, in
order to reduce the probability of error $f(\epsilon)$ by a factor of 10,
a reduction of $\epsilon$ by an order of $10^5$ is needed.
The improvement in prediction by increasing the accuracy of the
initial conditions is seen to be less favorable the smaller $\alpha$ is. 

Related to the above discussion, this paper presents a set of software
routines, implemented in MapleV R.4/5, together with an optional
interface in C for fast numerical integration, which, among other things,
calculates the
dimension of the fractal basin boundary.
The main goal is to determine numerically the
scaling law~(\ref{scaling}) and therefore the fractal dimension.
Consider an initial condition ${\bf X_0}$ and perturb one of the
coordinates, say $X_k$, in order to obtain two initial conditions
${\bf X} - {\bf \epsilon}$ and ${\bf X} + {\bf \epsilon}$.
By integrating
numerically we can determine to which attractor or asymptotic state both
of the perturbed initial conditions goes. We call the original initial
condition uncertain if the perturbed initial conditions have distinct
attractors. If a large number of initial conditions are chosen
randomly in a given region of the phase space, it is possible to
determine the fraction $\overline{f}(\epsilon)$ of these that are
uncertain, where  $\overline{f}(\epsilon)$ is 
the ratio between the number of uncertain conditions and the
total number of initial conditions. This process can be repeated for
several values of $\epsilon$. According to Bleher~{\it et. al.}
in~\cite{ott},
the quantities $f(\epsilon)$ and $\overline{f}(\epsilon)$ are
proportional, so that the parameter $\alpha$ can be determined by the
scaling of $\overline{f}(\epsilon)$ with $\epsilon$. 

The package has the following features:
\begin{itemize}
\item  user-friendly software tools for fast numerical 
integration of both Hamiltonian and non-Hamiltonian dynamical systems,
as well as a set of commands for the flexible presentation of the results;
\item an optional C interface for fast high-precision numerical integration;
\item tools for determining sets of initial conditions that satisfy
given constraints, expressed, e.g., as $g(X_1,X_2,\ldots,X_d) \geq 0$, 
where $g(X_1,X_2,\ldots,X_d)$ is a given function of $X_i$, $i=1,\ldots,d$;
\item the ability to perform combined symbolic and numerical studies by
implementing these software tools in a symbolic computing environment.
\end{itemize}

\noindent 
The paper is organized as follows: in
section~\ref{package} we give a summary of the package's commands, followed
by a detailed description\footnote{Aside from this, the package itself
contains on-line help in standard Maple format which can be
viewed as the user's manual for all the routines.} of its most relevant
commands, mainly {\tt Nsolve}, for the generation of initial conditions
and numerical integration of the system allowing 2D or 3D plots,
{\tt View}, for changing the ranges of all variables (including the
time) and {\tt Fdimension}, to determine the dimension of the fractal
basin boundary; section~\ref{examples} illustrates the application of
various commands
of the package to two examples with known results already presented in
the literature; section~\ref{perform} analyzes the performance of the various
integration schemes, and section~\ref{conclude} contains some concluding
remarks.

\section{The {\it Ndynamics} package}
\label{package}

Basically the {\it Ndynamics} package consists of a set of routines for
calculating the numerical evolution of dynamical systems and plotting
the results. The package generates initial conditions in a user-defined
region, subject to constraints provided by the user.  After calculating
the evolution of each trajectory, the user may use its plotting commands
to locate possible regions of interest --- regions where the system
presents chaotic behavior, and so on. Finally, a set of routines is
available to calculate the fractal dimension of the boundaries using
box-counting algorithms.

\subsection*{\it Summary}

A brief review of the commands of the package is as
follows:\footnote{This subsection and the next one may contain some
information already presented in the previous sections; this is
necessary to produce a self-contained description of the package.}

\begin{itemize}

\item {\tt Nsolve} takes as arguments a dynamical system, the {\em ranges\/}
of the initial conditions (from which it generates the initial conditions),
the range for the time
(independent variable), and the time step to be used when plotting
the associated trajectories.
The variables to be included in the final plot (which can be 2D or 3D)
must also be defined;

\item {\tt View} generates a plot from points already calculated,
allows the user to modify the ranges for all the variables (including
the time), and to swap between 2D and 3D plots;

\item {\tt Boxcount} applies the box counting algorithm to determine
the number of hypercubes of side $\epsilon$ (see the introduction)
covering the boundary;

\item {\tt Fdimension} uses {\tt Boxcount} for many values of
$\epsilon$ to calculate the fractal dimension of the boundary.

\end{itemize}

\subsection*{\it Description}

A complete description of the {\it Ndynamics} package's commands is
found in the on-line help. Here we present the most relevant part of
that description.

\subsection{Command name: {\tt Nsolve}}
\label{Nsolve}

\noindent {\it Feature:} plot 2D/3D graphs calculated for a set of initial
conditions, generated by the program, within ranges defined by the user.

\bigskip

\noindent
{\it Calling sequence\footnote{In what follows, the {\it input}
can be recognized by the Maple prompt \verb->-.}:}
\begin{verbatim}
> Nsolve(sys,[dep_ranges,[indep_range,plotstep]],Frame,
  optional_parameters);
\end{verbatim}

\noindent
{\it Parameters:}

\noindent
\begin{tabular}{ll}
\verb-sys-           & - a set of first order ordinary differential equations.
\\
\verb-dep_ranges- 
&  - a list with the ranges for the dependent variables in the format: \\
& [x=a..b, y=c..d, ...], where {x, y} are the dependent variables and \\
&  a..b, c..d, are the ranges in which the initial conditions will be \\
&  generated.  \\
\verb-indep_range-
& - the range for the independent variable: t=t1..t2, where t represents\\
& the independent variable.
\\
\verb-plotstep-      & - a numerical value defining the interval (in the
                        independent variable) \\ & for the plot.

\\
\verb-Frame-         & - a list with the variables to be plotted.
\end{tabular}

\bigskip

\noindent
{\it Optional Parameters:}
\smallskip
\smallskip

\noindent
\begin{tabular}{ll}
\verb-initial-
   & - Forces the command to stop after generating the initial \\
   & conditions.  \\
\verb-random-
   & - Tells the program to generate the initial conditions \\
   & randomly. \\
\verb-diagonal-
   & - Makes the program look for the initial conditions \\
   &   along the main diagonal of the hypercuboid defined by \\
   &  the ranges of the dependent variables. \\
\verb-method=option-
  & - Indicates which numerical method of integration will \\
  & be used, where \verb-option- can be any one of the options \\
  & available to dsolve in MAPLE (see the associated help).  \\
\verb-method=[rk5C,number]-
& - Tells the program to generate the C code and manage the\\
& interface to the C routines, which use a 5th order Runge-\\
& Kutta method to integrate the equations more rapidly. The \\
& parameter \verb-number- defines the integration step (constant).

\end{tabular}

\noindent

{\it Synopsis:}
\smallskip
\smallskip

The {\tt Nsolve} command is a part of the {\bf Ndynamics} package,
which, to brutally summarize things, is designed to calculate the
fractal dimension of boundaries. The command that actually does this
is {\tt Fdimension}.  In order to calculate this dimension, the process
involves the demarcation of dynamically interesting regions of the system
being studied.  This demarcation is achieved by running {\tt Nsolve} and
looking for such regions by analyzing the resulting graphs.  To facilitate
this, {\tt Nsolve} is designed to calculate and plot the trajectories
associated with a given system of differential equations, subject to
constraints and adjustments which can be tailored to the specific search
at hand.  In order to control and customize the output of {\tt Nsolve},
we have introduced the global and environment variables described below.

\begin{itemize}

\item{Global Variables}

\begin{enumerate}
\item {\bf Coloring}
\item {\bf constraint}    
\item {\bf initial\_conditions}
\item {\bf number\_ic}
\item {\bf initial\_time}
\end{enumerate}

The first item above controls in which color a trajectory will be
displayed. {\bf Coloring} is a boolean condition that is applied
to the last calculated point of each trajectory so that different
trajectories are displayed in one of two colors, corresponding to
the cases where {\bf Coloring} is true or false. {\bf Coloring}
defines the boundary, since it is defined to discriminate trajectories on
either side of it. If {\bf Coloring} is not assigned, then the command
uses a default setting, which produces a plot with a certain color
pattern. This is useful when the criteria to define the boundary have
not yet been decided, but it is desirable, perhaps for the purposes of
defining these criteria, to see the general flow of the trajectories.

The second item, {\bf constraint}, is a variable that defines the
constraints imposed upon the initial conditions. All initial
conditions generated will satisfy these constraints.
This variable can be any valid boolean Maple condition.
For instance, it may be two (or more) simultaneous conditions, such
as  $100 > x^2-y^2+z^4 > 20$, where $x,y,z$ are dependent
variables of the differential system. This variable does not have a
default value and can be left unassigned, in which case the command
freely  generates the initial conditions inside the hypervolume
defined by the dependent  variable ranges.

Item number 3 is a facility that allows the user to supply
the program with specific initial conditions. If this global variable is
assigned, the program skips the part that generates the initial
conditions and uses those supplied by the user. Its syntax
is the same as that produced by the program with
{\tt initial\_conditions} unassigned.  We refer the reader to the
on-line help that comes with the package for this syntax.

In the case that {\tt initial\_conditions} is not assigned by the user,
the global variable {\tt number\_ic} has to be defined, otherwise the
program issues an error message. As the name suggests, it defines the
number of initial conditions to be generated.

Finally, the global variable {\tt initial\_time} allows the user to define
the initial value for the independent variable to be anywhere in the
interval stipulated for its range\footnote{In the present version, when
using the C interface, this facility is not available.}.
The integration of the trajectory is then performed in both
directions, up to the limits of the time range.

\item{Environment Variables}

To customize the interface between the program and the user, the
following environment variables were introduced:

\begin{enumerate}
\item{\_Env\_print},
\item{\_Env\_store}.
\end{enumerate}

{\tt \_Env\_print} controls the level of printing during the execution of
the command. When set to zero, no printing takes place; if
{\tt \_Env\_print} =1, (the default value) basic messages are printed;
when set to 2, time-counting messages will be printed and,
finally, for the value 3 all possible messages are printed. 

The variable {\tt \_Env\_store}, controls in which file the results will
be stored. The initial conditions, the range for the variables, the
differential equation system, etc., are, by default, stored in a file
called usedata1 and the calculated points (trajectories) are stored (by
default) in a file called usedata2. This arrangement is fine if only
one process is running at a time. It will, however, result in file
conflicts if the user wishes to perform simultaneous runs of the program
from the same directory. The assignment of {\tt \_Env\_store} by the
user to the desired filenames avoids such problems.

\end{itemize}

\medskip
\noindent
{\it The arguments}
\smallskip
\smallskip

The first argument of {\tt Nsolve} is a set containing the system of
differential equations.
In the present implementation, the program only deals with
first order ordinary differential equations (ODEs).
This is not a major drawback, since all
systems of ODEs can be put into this form. The second argument
is a list in which the first element is itself a list giving the ranges of
the dependent variables. The second element is a list containing the range
of the independent variable and the step (for that variable) for printing the
graphs. The third argument gives the variables to be plotted (this can be two
or three variables). Finally there are the optional arguments. 

\bigskip

\noindent
{\it The optional arguments}
\label{Nsolveoptional}
\smallskip
\smallskip

The optional arguments can be given alone or in conjunction and in any
order.
By default, {\tt Nsolve}
takes care of everything for the user --- it
generates the initial conditions and fully integrates the system. If the
optional argument {\it initial} is given then it stops after calculating
(and saving in the appropriate file) the initial conditions.

The optional arguments {\it random} and {\it diagonal} act similarly.
There are thus three ways in which the command may generate the initial
conditions in the hypercuboid defined by the dependent variable ranges.
The default is to produce initial conditions homogeneously
throughout the hypercuboid.
However, if the argument {\it random} is used then the initial
conditions will be randomly distributed inside 
hypercuboid. If, instead, {\it diagonal} is used, all conditions will be
generated along the main diagonal
\footnote{By ``main diagonal'' we mean the straight line inside
the hypercuboid which joins the point defined by the minimum value
of the ranges of the independent variables to the point defined by
the maximum values of those ranges.}
of the region.

Many different algorithms can be used to numerically integrate the
system. Maple provides a great variety of these and all of them can be
used with our command, via the option {\it method}.
For example, if {\it method=classical[rk2]},
the second order Runge-Kutta method will be used
\footnote{Details on the possible numerical solving methods
available (and their syntax) are available via the MAPLE help
page for the command {\tt dsolve[numeric]}.}.


Apart from the numerical methods available in the MAPLE system, our
program has a further possibility: a C-based numerical
integrator. The only additional requirement for this is that the
machine where the program is being run has a C compiler.
In this case, the integrator is a fixed-step fifth order Runge-Kutta scheme
based on the procedure {\tt RKQC} in~\cite{NumRecC}.
The idea behind this option is to gain speed over Maple's interpreted
procedures, thereby combining the flexibility of the symbolic environment
with the efficiency of the compiled C routine (see section~\ref{perform}
for comparisons).
It is our intention to introduce more compilable algorithms in later releases.


\noindent
\subsection{Command name: {\tt View}}
\label{View}

\noindent {\it Feature:} This commands allows easy and efficient viewing of
different parts of the trajectories calculated by {\tt Nsolve},
allowing the user to zoom in and out, and to change the
variables being viewed (if necessary the user can even swap between
2D and 3D plots).
\smallskip
\smallskip
\smallskip

\noindent {\it Calling sequence:}
\begin{verbatim}
> View(list_range);
\end{verbatim}
\smallskip
\smallskip

\noindent
{\it Parameters:}
\smallskip
\smallskip

\noindent
\begin{tabular}{ll}
\verb-list_range- 
  & - a list with the new ranges of those (two or three) variables\\
  & the user wants to view in detail.  The format is: [x=a..b, y=c..d, ...],\\
  & where {x, y} are (dependent or independent) variables and a..b, c..d, \\
  & their ranges. 
\end{tabular}
\smallskip
\smallskip

\noindent
{\it Synopsis:}
\smallskip
\smallskip

The importance of this command lies in the search for the relevant
regions for studying the system's behavior. The user 
generally starts by running {\tt Nsolve} in a broad
region of the variable space and then studying more closely the regions,
for example, where chaos is suspected. The graphs that {\tt View}
produces (2D or 3D) do not need to display the same
variables that {\tt Nsolve} uses in its argument {\it Frame}.

\noindent
\subsection{Command name: {\tt Boxcount}}
\label{Boxcount}

\noindent 
{\it Feature:} this command analyzes the effect of a small perturbation on
the initial conditions on the fate of a bundle of trajectories.
\smallskip
\smallskip
\smallskip

\noindent {\it Calling sequence:}
\begin{verbatim}
> Boxcount(epsilon, final_time, optional_parameters);
\end{verbatim}

\noindent
{\it Parameters:}
\smallskip
\smallskip

\noindent
\begin{tabular}{ll}
\verb-epsilon-   & -  defines the size of the perturbation around each
                      initial condition. 
\\ 
\verb-final_time-   & -  indicates the final value of the independent
                    variable.

\end{tabular}

\bigskip

\noindent
{\it Optional Parameters:}
\smallskip
\smallskip

\noindent
\begin{tabular}{ll}

\verb-method=option-
& - Indicates which numerical method of integration will \\
& be used, where \verb-option- can be any one of the options \\
& available to dsolve in MAPLE (see the associated help). \\
\verb-method=[rk5C,number]-
& - Tells the program to generate the C code and manage the\\
& interface to the C routines, which use a 5th order Runge-\\
& Kutta method to integrate the equations more rapidly. The \\
& parameter \verb-number- defines the integration step (constant).
\end{tabular}

\medskip
\noindent
{\it Synopsis:}
\smallskip
\smallskip


Suppose we have generated a set of initial conditions
(see~\ref{Nsolveoptional}).
{\tt Boxcount} perturbs each of these initial conditions within a
neighborhood of radius = $\epsilon$.  For points sufficiently close to
the boundary (that separates the different end-point behaviors)
the two perturbed trajectories will have different fates.  {\tt Boxcount}
then counts the number of initial conditions for which the pairs
of perturbed trajectories evolve to different attractors.

\noindent
\subsection{Command name: {\tt Fdimension}}
\label{Fdimension}

\noindent {\it Feature:} Fundamentally applies {\tt Boxcount}
a user-defined number of times, with a different value of $\epsilon$
each time.
\smallskip
\smallskip
\smallskip

\noindent {\it Calling sequence:}
\begin{verbatim}
> Fdimension(epsilon_range, TF, Nepsilon, optional_parameters);
\end{verbatim}

\noindent
{\it Parameters:}
\smallskip
\smallskip

\noindent
\begin{tabular}{ll}
\verb-epsilon_range- 
&  -  defines the range of variation of $\epsilon$.  \\ 
\verb-TF-
& -  indicates the final value of the independent variable.  \\ 
\verb-Nepsilon-
& -  indicates how many values of $\epsilon$ will be considered in  \\
&  epsilon\_range.

\end{tabular}

\noindent
{\it Optional Parameters:}
\smallskip
\smallskip

\noindent
\begin{tabular}{ll}

\verb-savefile=filename-
  & - Indicates in which file the results will be stored. The  \\
  & default value of {\it filename} is {\tt XY-file}. Since \\
  & {\tt Fdimension} makes many calls to {\tt Boxcount}, only \\
  & the pairs of data $\epsilon$, and the logarithm of the \\
  & ratio of the number of initial conditions on the boundary \\
  & to the total number of initial conditions (see appendix~\ref{frac}) \\
  & are saved to the file.\\

\verb-method=option-
&  - Indicates which numerical method of integration will \\
& be used, where \verb-option- can be any one of the options \\
& available to dsolve in MAPLE (see the associated help). 
\\

\verb-method=[rk5C,number]-
& - Tells the program to generate the C code and manage the\\
& interface to the C routines, which use a 5th order Runge-\\
& Kutta method to integrate the equations more rapidly. The \\
& parameter \verb-number- defines the integration step (constant).

\end{tabular}
\smallskip
\smallskip
\smallskip
\smallskip

\noindent
{\it Synopsis:}
\smallskip
\smallskip


{\tt Fdimension} essentially implements the box-counting process in
our program. It uses {\tt Boxcount} to evaluate the number of
points on the boundary for many values of $\epsilon$ thus
calculating the fractal dimension of the boundary
(using the method described in appendix~\ref{frac}).

\section{Examples}
\label{examples}

To illustrate the routines presented in the previous section we consider
the well-known Lorenz system\cite{lorenz}:

\begin{eqnarray}
\label{lll}
\dot{x} & = & \sigma (y-x), \nonumber \\
\dot{y} & = & - y - x z + R x,  \; \\
\dot{z} & = & x y - b z, \nonumber
\end{eqnarray}

\noindent where $\sigma$, $R$ and $b$ are constant parameters of the problem.
$R$ plays a crucial role as far as the dynamics of phase space
is concerned. A simple analysis shows the presence of three critical
points (the points satisfying the conditions $\dot x=\dot y=\dot z =0$):
the origin $P_0(0,0,0)$ and, provided $R>1$,
two further points are symmetrical with respect to the $z$-axis,
$P_{\pm}=(\pm \sqrt{b (R-1)},\pm\sqrt{b (R-1)},R-1)$. The
stability properties of these critical points depend only on $R$, so
that we set \cite{lorenz} $\sigma = 10$ and $b = 8/3$. We summarize the
behavior of the solutions of the Lorenz system for distinct values of
$R>1$ that will be important in the applications of the routines and
commands we have described. For $1<R<24.74$ the origin $P_0$ is
unstable and $P_{\pm}$ are attracting stationary solutions, therefore
representing two possible asymptotic configurations (except for the set
of trajectories of measure zero that stay in the neighborhood of $P_0$).
However, around the critical value, $R_c \approx 13.926$, $P_0$ 
develops into a homoclinic
point such that, above this value, the basins of attraction around
$P_-$ and $P_+$ are no longer distinct. This means that trajectories can
cross backwards and forwards between the two before settling down. Another
transition occurs when $R \approx 24.74$, for which $P_-$ and $P_+$ become
unstable. Lorenz considered $R=28$ and obtained a remarkable behavior
displaying for the first time a strange attractor. As we shall show,
the dynamics are highly erratic: a given
trajectory can spiral around one of the critical points, $P_+$ or $P_-$,
for some arbitrary period then jump to the neighborhood of
the other critical point, spiral around that for a while and then
jump back to the first one, and so on. 

In Figure~1 we show the form of the trajectories for the Lorenz system,
with coefficients $\sigma = 10$, $b = 8/3$ and $R=20.06$. 
We are therefore inside the parameter range for erratic behavior discussed
above.  We see that, given enough time (in our case $\gtrsim\,50$
units of time),
the trajectories settle down around one of the two possible asymptotic points, 
$P_{\pm}$.
Figure~1 was obtained with 10 initial conditions,
$(x_0,y_0,z_0)$, that were randomly chosen in the volume of phase
space defined by $0.9 < x_0 < 1.1$, $0.9 < y_0 < 1.1$
and $21.9 < z_0 < 22.1$.
This is our first illustration of the application of 
{\tt Nsolve}. Figure~1 was obtained using the set of commands below.
First, we set the accuracy, read the program and define the system:

\begin{verbatim}
> Digits := 16;
\end{verbatim}
\begin{verbatim}
> with(Ndynamics);
\end{verbatim}
\begin{tt}
\small
\centerline{[Boxcount, Fdimension, Nsolve, View]}
\end{tt}
\begin{verbatim}
> sigma:=10.0: b:=8.0/3: R:=20.06:
\end{verbatim}
\begin{verbatim}
> Lorenz:={diff(x(t),t)=sigma*(y(t)-x(t)),
           diff(y(t),t)=-x(t)*z(t)+ R*x(t)-y(t),
           diff(z(t),t)=x(t)*y(t)-b*z(t)};
\end{verbatim}

Note that the integers in (\ref{lll}) have been swapped for
floating-point numbers. This is important in the case of the
parameter $b$ to avoid truncated integer division occurring in
the C subroutines generated.

We then define the input:

\begin{verbatim}
> x_range := 0.9..1.1: y_range := 0.9..1.1: z_range := 21.9..22.1:
> t_range := 0..50:
> plot_spacing := 0.03:
> Frame := [x(t),y(t),z(t)]:
> number_ic := 10:
> time_step := 0.001:
\end{verbatim}

Now we can call {\tt Nsolve}:

\begin{verbatim}
> graph := Nsolve(Lorenz,[[x=x_range,y=y_range,z=z_range],
           [t=t_range,plot_spacing]],Frame,method=[rk5C,time_step]):
\end{verbatim}

\vspace{0.5cm}

\noindent

The use of {\tt Nsolve} above illustrates the application of
the global variable {\it Coloring}. Here we have based the choice of
color by specifying the approximate location ($x$ coordinate) of one of
the asymptotic points $P_{\pm}$. Trajectories ending around this point are
colored green, while trajectories ending around the other critical point are
colored black. Using {\tt Display(graph)} gives the result shown in Figure~1. 

In this paper we take as a measure of the degree of chaos of a dynamical
system, such as the Lorenz system above, the fractal dimension
associated with the possible different exit modes under small changes of
initial conditions, as discussed in section~\ref{intro}.  A note here is
in order
about our criterion for choosing the intervals between the values of
$\epsilon$ (the variations for the initial conditions as explained 
in section \ref{Fdimension}). Given ranges for the initial
conditions which define a hypervolume $V_{\rm ic}$ in $d$ dimensions, and
distributing $N_{\rm ic}$ initial conditions throughout this volume, the
largest $\epsilon$ cannot exceed the average distance $\bar{l}$ between
the initial conditions, since otherwise the ``perturbation'' $\epsilon$
for different initial conditions will overlap leading to loss of
predictability of the fractal dimension. We take for $\bar{l}$ the
minimum distance between two initial conditions for a uniform
distribution, which is easily deduced to be

\begin{equation}
\bar{l} = \frac{V_{\rm ic}^{1/d}}{N_{\rm ic}^{1/d} -1} \;.
\label{average-l}
\end{equation}

\noindent 
In all the numerical tests performed the largest value for $\epsilon$
is of the order of $\bar{l}$. The lowest value for $\epsilon$ used is
determined by the statistical error associated with the number of initial
conditions $\bar{N}$ falling inside the fractal boundary. We then have that
$\bar{N}^{1/2}$ will give the statistical uncertainty associated with the
probability $f = \bar{N}/N_{\rm ic}$ of a given initial condition
falling within the boundary \cite{ott}. The numerical tests performed
have values of $\epsilon$ and $N_{\rm ic}$ such that
$\bar{N} \, \gtrsim \, 400$,
thus assuring an estimated error ($\sqrt{N}/N$) of at most $5 \%$ in
the determination of $f$. 

The
fractal dimension can be readily obtained by using the command {\tt
Fdimension}. We begin by generating the initial conditions as explained in
section \ref{Nsolve}:

\begin{verbatim}
> number_ic := 10000:
> Nsolve(Lorenz,[[x=x_range,y=y_range,z=z_range],
         [t=t_range,plotting_spacing]],Frame,initial):
\end{verbatim}

\vspace{0.5cm}

\noindent
Note that 10000 initial conditions are used. By omitting the optional
parameter {\it random} (see section \ref{Nsolve}), these $number\verb3_3ic$ initial
conditions are uniformly distributed throughout the interval volume 
$V_{\rm ic}=(0.2)^3$ defined by the dependent variable ranges.
The use of random initial conditions
leads to no appreciable change in the results, which always remain
within the statistical error defined above. 
Values of $\epsilon$ ranging from $10^{-5}$ to $10^{-4}$ were used
in {\tt Fdimension} when computing the fractal dimension.
An easy way to check the number of initial conditions, $\bar{N}$,
which are generated inside the basin boundary
is by using {\tt Boxcount}. For example, having generated 10000 initial 
conditions, we can use:

\begin{verbatim}
> Boxcount(0.00001, 60, method=[rk5C,0.01]);
\end{verbatim}
\begin{tt}
\small
\centerline{From the,10000, points (that were testable),2202, were close to
the boundary.}
\end{tt}
\begin{tt}
\centerline{[2202,10000]}
\end{tt}

\bigskip

\noindent
tells us that, for the $N_{\rm ic} = 10000$ initial conditions and for
$\epsilon = 10^{-5}$, $\bar{N} = 2202$ initial conditions are around
the boundary, with a resulting statistical error of around 2\%.
If we decrease $\epsilon$, say to $10^{-10}$,  we get instead
$\bar{N} = 551$, and the statistical error is of the order
of $N^{-1/2}\approx 4\%$, close to the 5\% limit of acceptability
which we have imposed. Accordingly, for smaller values of $\epsilon$
we must increase the number of initial conditions in order to
assure that we have at least the minimum number of initial conditions
on the boundary necesary for statistical acceptability.
With this done, the fractal
dimension is then obtained by issuing the command:

\begin{verbatim}
> Fdimension(0.00001..0.0001, 60, 10, method=[rk5C,0.01]);
\end{verbatim}

\noindent 
where we are evolving the trajectories through $60$ units of time and 
we are taking $10$ equally spaced points in the range
$\ln (10^{-5})$ to $\ln (10^{-4})$. The results obtained are shown in
Figure~2 in terms of the logarithm of the number of uncertain initial
conditions, (those that change their fate, or ``color'' on the graph),
against $\ln(\epsilon)$. The slope of the best-fit straight line through
the points gives the number $\alpha$ (see Appendix).
The fractal dimension is then defined from~(\ref{D_0}) as
$D_0=d-\alpha$. We obtain for the fractal dimension the result
$D_0 \simeq 2.87 \pm 0.01$, in complete agreement~\cite{lor-res} with the
known result for the Lorenz system for the choice of parameters
$\sigma = 10$, $b = 8/3$ and $R=20.06$.

In Figure~3 we once again compute the fractal dimension with
{\tt Fdimension}, but with
$10^{-8}\leq\epsilon\leq10^{-7}$. The result is the same 
up to statistical error.

As a further check of our set of routines, we evaluate the fractal
dimension for the set of parameters $\sigma = 10$, $b = 8/3$ and $R=10$,
in the parameter region prior to the onset of transient
chaos~\cite{caos2}. Using {\tt Fdimension}, with
$5\times 10^{-3}\leq\epsilon\leq 9\times 10^{-3}$ (an interval that
ensures the minimum statistical error), we obtain the result shown in
Figure~4, from which we conclude that $D_0 \simeq 1.98 \pm 0.05$.
Further refinements of this result for a larger number of initial
conditions leads us to conclude that $D_0 = 2$ and $\alpha = 1$ and
that, as expected, there is no chaos for this choice of parameters.

As a demonstration of the use of {\tt View}, we
illustrate the basin-boundary in the two cases analyzed above.
Figure~5 shows the case $R=10.00$ while Figure~6 illustrates the
results for $R=20.06$. In both cases $number\_ic$ is taken to be 100.

The arguments to {\tt View} are $[x=-0.1..0.1,y=-0.1..0.1,z=0.9..1.1]$ (for Figure~5)
 and $[x=0.9..1.1,y=0.9..1.1,z=21.9..22.1]$ (for Figure~6). 
From Figure~5 we easily identify a well-defined (non-fractal) boundary
separating the sets of initial conditions whose associated trajectories
approach $P_+$ and $P_-$.
In contrast we see in Figure~6 that neighboring
initial conditions may lead to very different end-states.
This is the fractal basin boundary discussed in the introduction.

\section{Performance}
\label{perform}


To point out the advantages of our hybrid symbolic/numeric approach,
which uses Maple to manage the source code generation and compilation
for the number crunching, while maintaining Maple's flexibility,
table~\ref{perftab} presents a comparison of the elapsed time taken
to perform the same calculation using the C interface and performing
the entire calculation in Maple, using some of its numerical
integration routines.  To obtain a fair comparison particular care
was taken to use methods of the same order and adjusting parameters
to produce solutions with the same precision.
The following procedure was used: for a given set of initial conditions, namely
$x_0=0.679149319354506$, $y_0=-0.5692394267519960$ , $z_0=22.00017552807044)$,
Maple's inbuilt high-precision Taylor Series intergrator was used to integrate
system~(\ref{lll}) from $t=0$ to $t=11$ until
converegence was obtained to an accuracy of 13 decimal places;
subsequently Maple's
global variable {\tt Digits} was set equal to 13 (higher values of {\tt
Digits} seemed to cause problems with Maple's integrator), and Maple's inbuilt
5th order Runge-Kutta-Fehlberg integrator, {\tt rkf45}, was used up to $t=11$.
This integration was found to be accurate to 4 decimal places. The
step size in the C routine was then adjusted to give the same\footnote{in
fact, the C routine was slightly more precise} precision, resulting in
a step size of 0.002. An average was then taken for 200 integrations.

The results below were obtained with Maple~V.4 running in Windows~98
on an AMD-K6 266 with 64Mb of SDRAM. The C compiler used was
Delorie's implementation of GNU's gcc, available form www.delorie.com,
with level~3 optimization.


\begin{table}[htb]
\begin{center}
\begin{tabular}{|c|c|c|}
\hline 
{\bf Integration}  & {\bf C-interface} & {\bf Maple} \\
{\bf Method}  & {\bf rk5} &{\bf rkf45} \\
\hline 
Seconds per Trajectory  & 0.1 & 7.0 \\
\hline
Ratio to the Fastest Case  & 1 & 70 \\
\hline 
\end{tabular}
\caption{Comparative performances of Maple's inbuilt numerical integrators
and our C interface for the Lorenz system.}
\label{perftab}
\end{center}
\end{table}

\section{Conclusions}
\label{conclude}

We have presented a set of software tools which allow great flexibility
for the analysis of dynamical systems composed of first-order ordinary 
differential equations, Even though the construction
of these programs was mainly motivated by the computation
of fractal dimensions of basin boundaries of chaotic systems, this is not
their only use, as was highlighted in the previous section. Large regions
of phase space can be readily studied with a fairly large number of
initial conditions, allowing the user to quickly find regions
of interesting behavior. 

The optional numerical interface allows for faster numerical
integration than with Maple's internal routines. The ease of
use of the Maple software is thus combined with a faster numerical
interface (here implemented in C, though this can be easily modified to
use routines written in other languages such as Fortran),
thereby allowing intensive numerical study of dynamical systems.


\begin{appendix}
\section{Fractals}
\label{frac}

\subsection{Fractal Dimension}


The idea of a fractal is hard to define precisely. But,
when we refer to a set {\it F} as a fractal, we typically have
the following in mind:

\begin{enumerate}
\item {\it F} has a ``fine structure'', i.e., detail on arbitrarily small
scales;
\item {\it F} is too irregular to be described in traditional
geometrical language, both locally and globally;
\item often {\it F} has some form of self-similarity, perhaps
approximate or statistical;
\item usually the ``fractal dimension'' of {\it F} (defined in some way)
is greater than its ``topological dimension'';
\item in most cases of interest, {\it F} is defined in a very simple way,
perhaps recursively.
\end{enumerate}


There are many ways to define the dimension of a fractal. An
important one is the {\it Hausdorff dimension}~\cite{haus} --- one possible
generalization of the ``primitive'' notion of dimension.

Suppose that we have a hypercube of side {\it a}. Its hypervolume is

\begin{equation}
\label{dim1}
H_v = a^d,
\end{equation}

\noindent
where $d$ is the dimension of the hyperspace.

If we divide the hypervolume into $N$ hypercubic cells of side $\epsilon$,
we have that

\begin{equation}
\label{dim2}
H_v = N \epsilon^d.
\end{equation}

\noindent
Dividing~(\ref{dim2}) by~(\ref{dim1}) we obtain
\begin{equation}
\label{dim3}
1 = N { \left( {\frac{\epsilon}{a}} \right) }^d.
\end{equation}
\noindent
Defining $\delta\equiv {\epsilon}/{a}$ and noticing that the number of
cells is a function of $\delta$ ($N = n(\delta)$)

\begin{equation}
\label{dim4}
1 = n(\delta) \, \delta^d.
\end{equation}

\noindent
Taking logarithms of both sides of~(\ref{dim4}) and solving for $d$, we
obtain the standard definition of dimension:
\begin{equation}
\label{dim5}
d = - {\frac{ \ln(n(\delta))}{\ln( \delta )}}
\end{equation}

\noindent
Suppose now that we want to measure the dimension of a Cantor set,
constructed as follows (see figure):

We take each segment (figure 7A), divide it into three equal parts, 
remove the middle segment and replace it with two other
segments of the same length (figure 7B), repeating this procedure {\it ad
infinitum} (figure 7C, ..), one gets (figure 7D).

To cover the Cantor set (figure tal n) with segments it would be necessary to
use an infinitesimal $\epsilon$, implying that $\delta \rightarrow 0$.
Hence the analogue of equation (\ref{dim5}) would be

\begin{equation}
\label{dim6}
d=-\lim_{\delta \rightarrow 0}
       {\frac{ \ln(n(\delta))}{\ln \left( \delta \right) }},
\end{equation}
\noindent
which is the Hausdorff dimension.

\noindent
In the Cantor set above
we can easily determine the value for the limit on~(\ref{dim6}):
\begin{equation}
\label{dim7}
d = - {\frac{ \ln(4)}{\ln \left( {\frac{1}{3}} \right) }} \sim 1.26.
\end{equation}

\subsection{Fractal Dimension of Boundaries}
\label{dimbound}

Consider a $d$-dimensional finite region $R$, and a subset $S\subset R$.
How can we determine the fractal dimension of $S$, given some
criteria as to whether a point of $R$ belongs to $S$ or not?


One possible way to determine the fractal dimension of $S$ is as follows.
Without loss of generality, let us suppose that $R$ is a hypercube
of side $a$.
Dividing $R$ into $N$ hypercubic ``cells'' of side $\epsilon$, and
evaluating the number of cells
$N_s$
that contain points belonging to the subset $S$
we expect that for $\delta \ll 1$
we have a good approximation to the
exact result given in~(\ref{dim6}).


However the number of cells 
\begin{equation}
N = \delta^{-d}
\end{equation}
rapidly increases
as $\delta \rightarrow 0$, and this approach becomes impractical. An
alternative approach is that of selecting $N^{*}$ random cells in $R$
and counting the number of 
cells $N^{*}_s$ that have points belonging to $S$. We expect that
\begin{equation}
\frac{N^{*}_s}{N^{*}} \rightarrow \frac{N_s}{N}
\end{equation}
as $N^{*}$ becomes statistically sound.
Supposing that, indeed,
\begin{equation}
\frac{N^{*}_s}{N^{*}} = \frac{N_s}{N}
\end{equation}
we have
\begin{equation}
\underbrace{- \frac{\ln(N_s)}{\ln \delta}}_{d_f} =
- \frac{\ln({N^{*}_s}/{N^{*}})}{\ln \delta} 
\underbrace{- \frac{\ln(N)}{\ln \delta}}_{d}.
\end{equation}
Defining
\begin{equation}
\alpha \equiv d - d_f = \frac{\ln(N^{*}_s/ N^{*})}{\ln \delta},
\end{equation}
we have
\begin{equation}
\ln \left( \frac{N^{*}_s}{N^{*}} \right) = \alpha \ \ln \delta.
\end{equation}
Both $\delta$ and $N^{*}_s/N^{*}$ can be measured, allowing us
to determine $\alpha$ and the fractal dimension $d_f$ of $S$.


In the previous discussion, little has been said about how to determine
whether or not a point belongs to the subset $S$. In our specific case
of differential equations, the following criterion is used. For a given
cell, a number of points within the cell is chosen, and the trajectories
passing through those points are followed. If these trajectories have
different attractors, the cell is said to belong to $S$.

\end{appendix}

\vspace{1.0cm} 
\begin{center}
{\large \bf Acknowledgments}
\end{center}

\vspace{0.5cm}

HPO and ROR are partially supported by CNPq. 
ROR also acknowledges partial support from FAPERJ.

\newpage

\begin{center}
{\large \bf Figure Captions}
\end{center}

\vspace{1.cm}

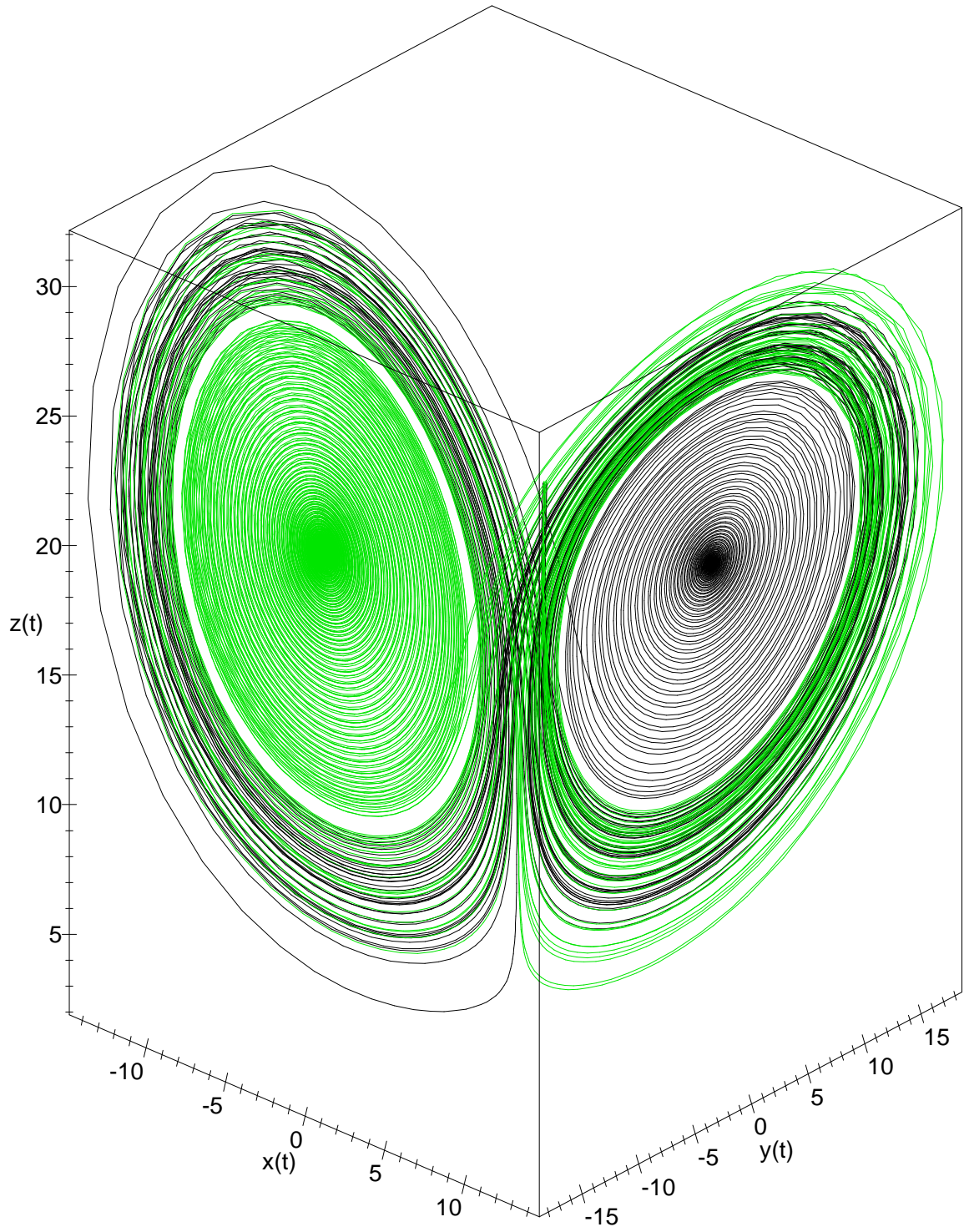
{\bf Figure 1:} A plot of the Lorenz system~(\ref{lll}) using
                {\tt Nsolve}.

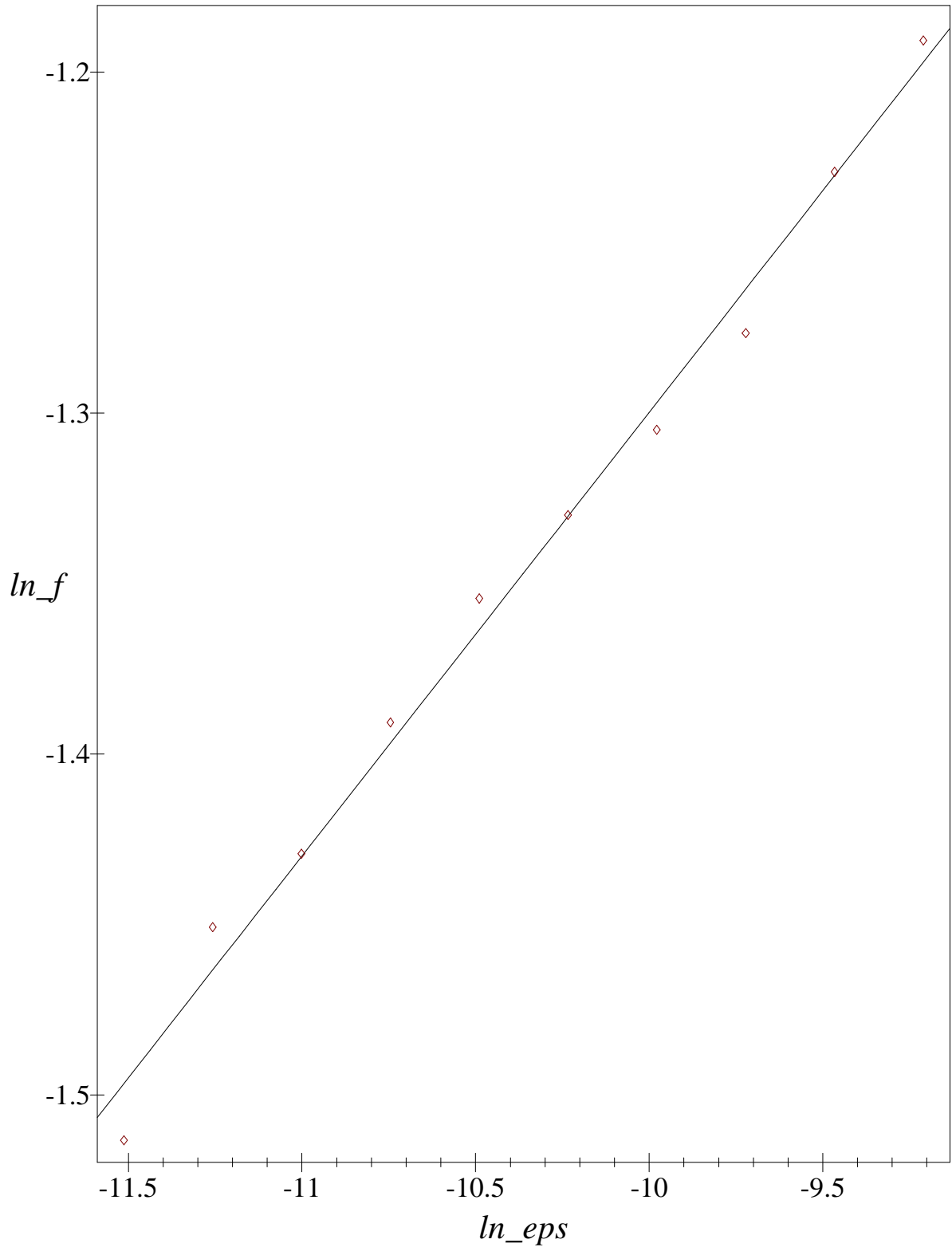
{\bf Figure 2:} $\ln(\bar{N}/N_{\rm ic})$ as a function of $\ln \epsilon$.
$N_{\rm ic} =10^4$ and $10^-5 \leq \epsilon \leq 10^{-4}$.
The slope of the curve, $0.13 \pm 0.01$, is the exponent $\alpha$
in~(\ref{D_0}) defining the fractal dimension: $D=2.87 \pm 0.01$.

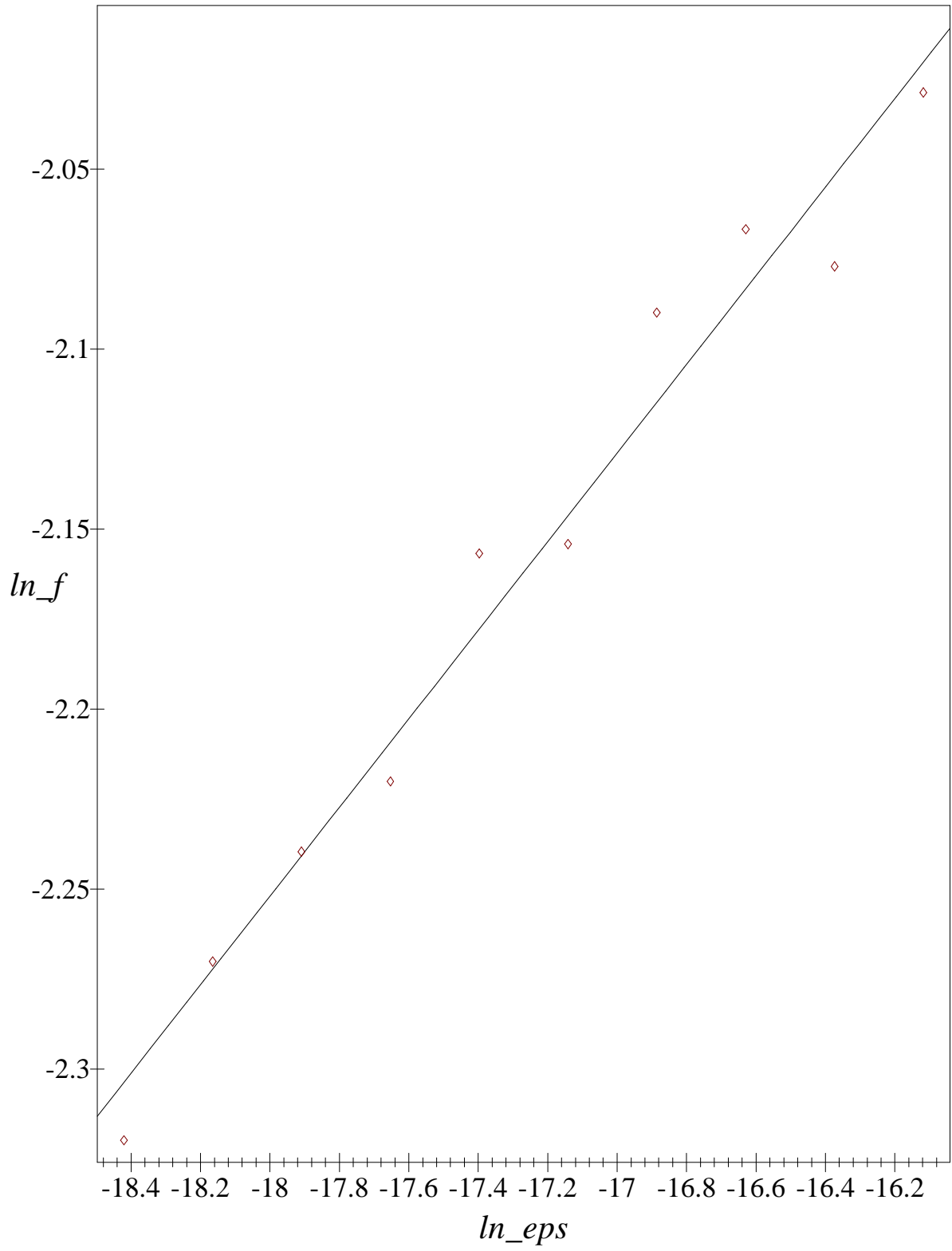
{\bf Figure 3:}As for Figure~2, but with $10^{-8} \leq \epsilon \leq 10^{-7}$.
The slope of the best-fit straight line is $\alpha = 0.12 \pm 0.01$.

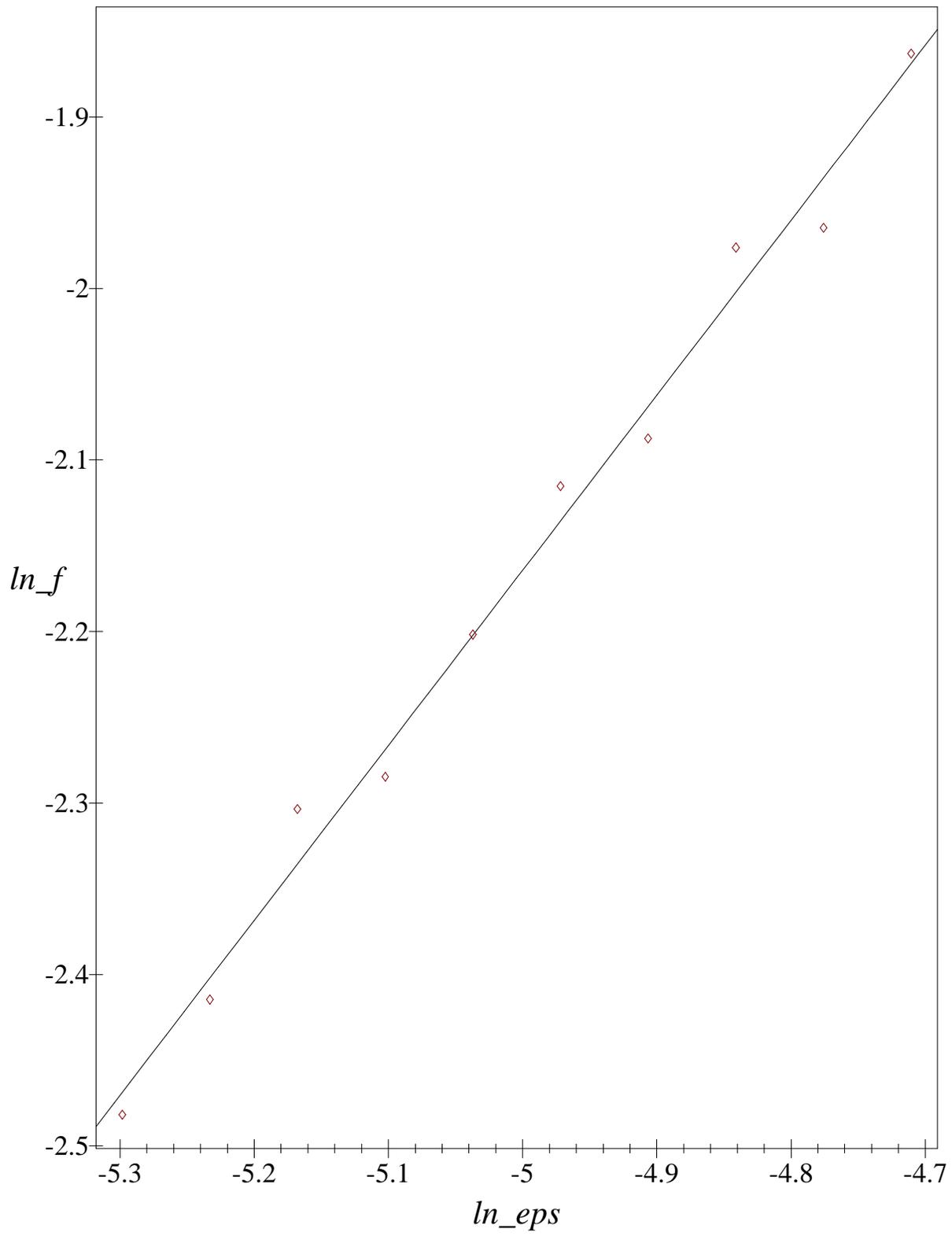
{\bf Figure 4:}The output from {\tt Fdimension} applied to the
non-chaotic (non-fractal) Lorenz system ($\sigma = 10$, $b = 8/3$ and $R=10$),
with $5\times 10^-3 \leq \epsilon \leq 9\times 10^{-3}$.
The slope of the best-fit straight line is $\alpha = 1.02 \pm 0.05$.

{\bf Figure 5:}An example of the use of {\tt View} for a
non-fractal system. 100 (random) initial conditions have been taken in
the phase space interval $[x=-0.1..0.1,y=-0.1..0.1,z=0.9..1.1]$.

{\bf Figure 6:}The result of 100 (random) initial conditions taken in the phase
space interval $[x=0.9..1.1,y=0.9..1.1,z=21.9..22.1]$, for a chaotic regime. 

{\bf Figure 7:}The construction process for the Cantor Set



%
%
%
%
%
%
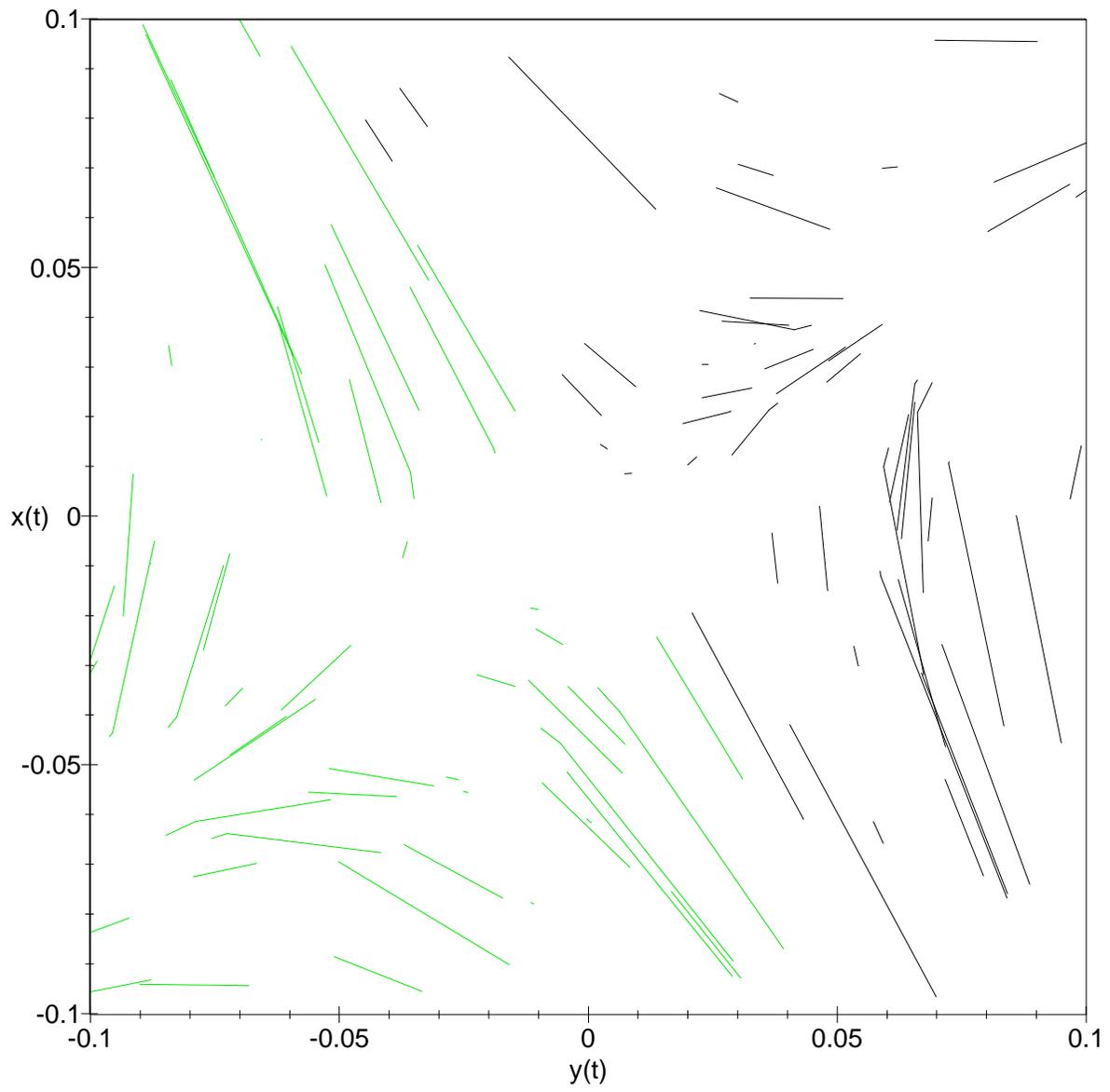
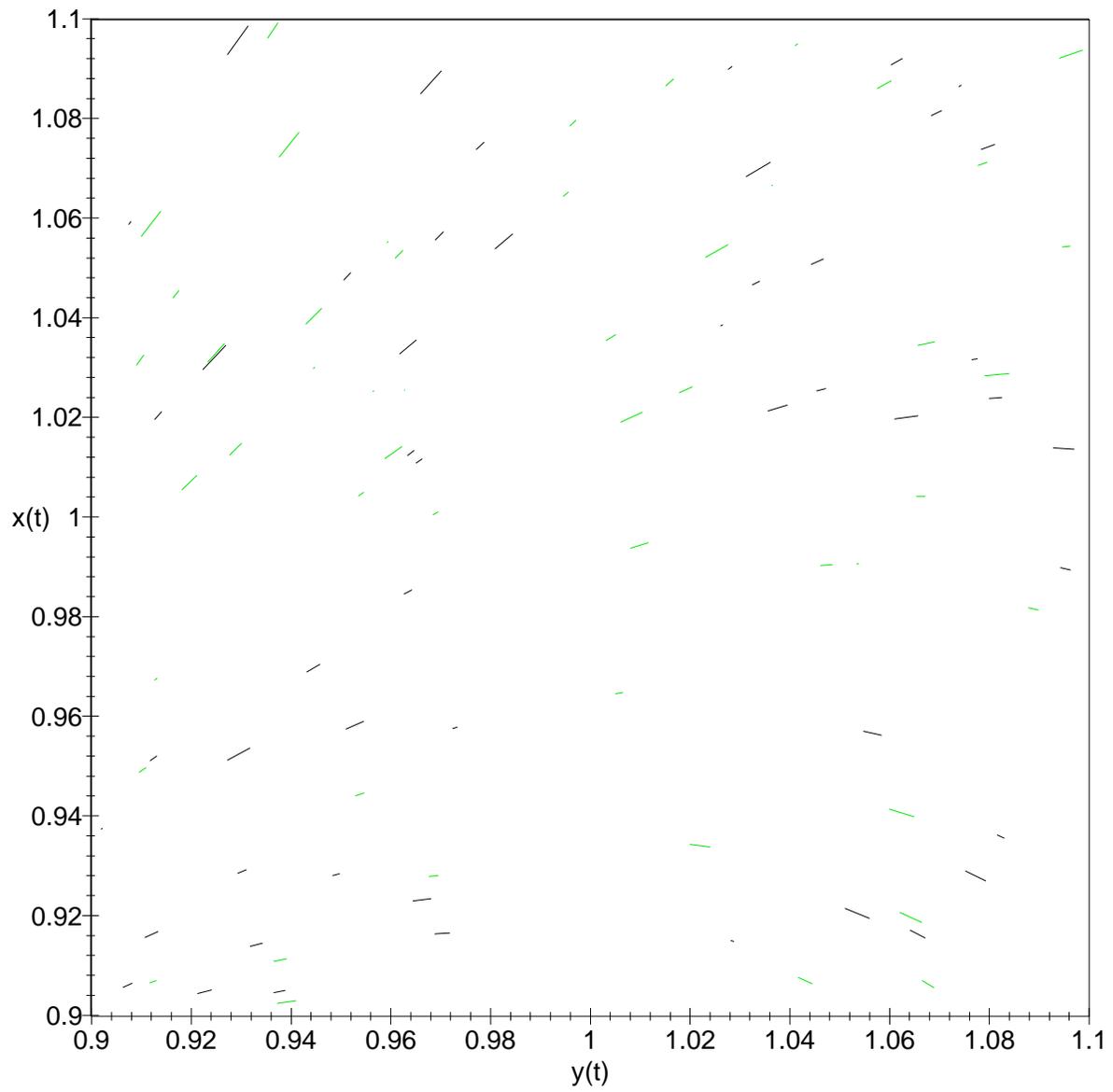
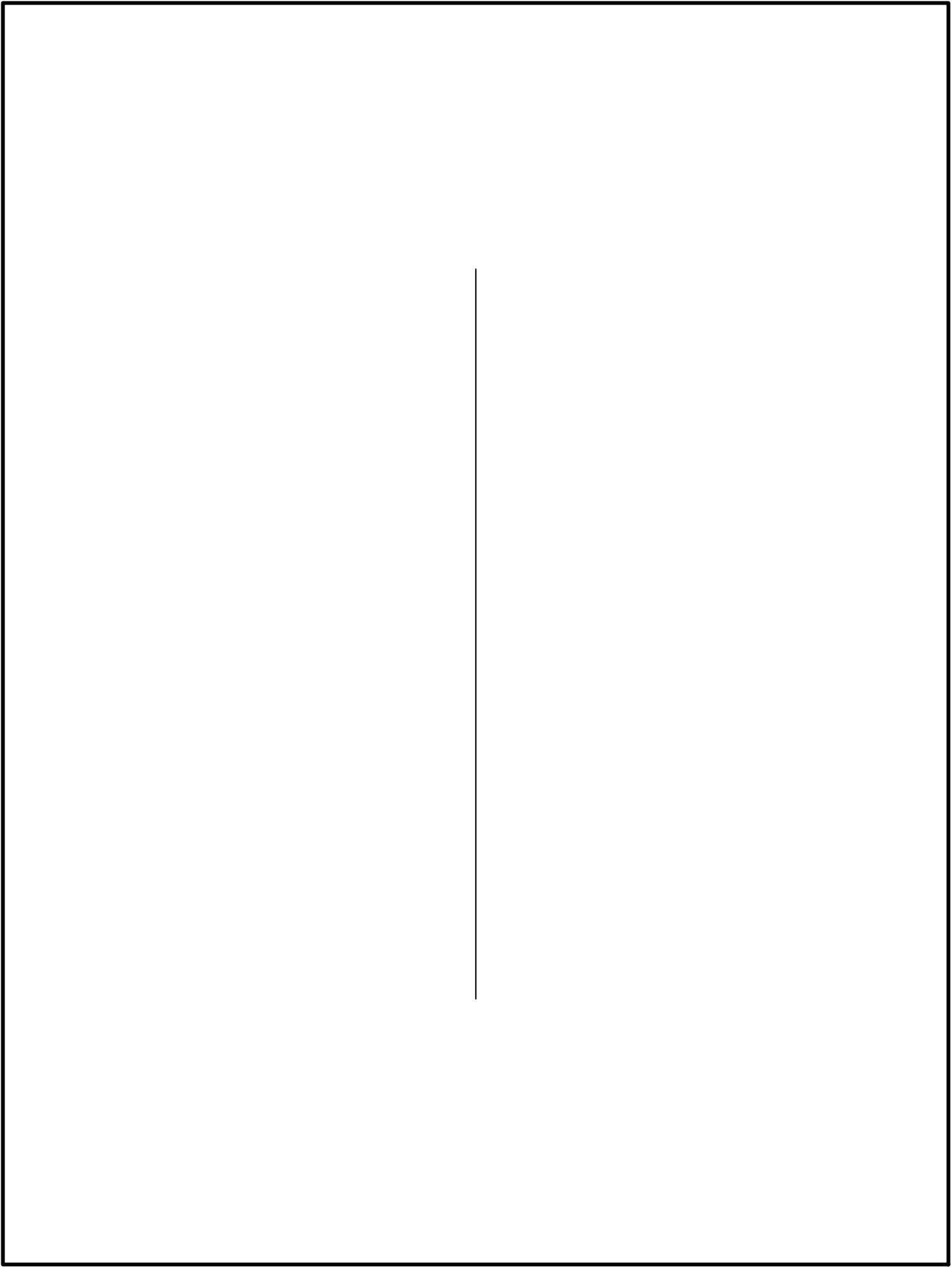
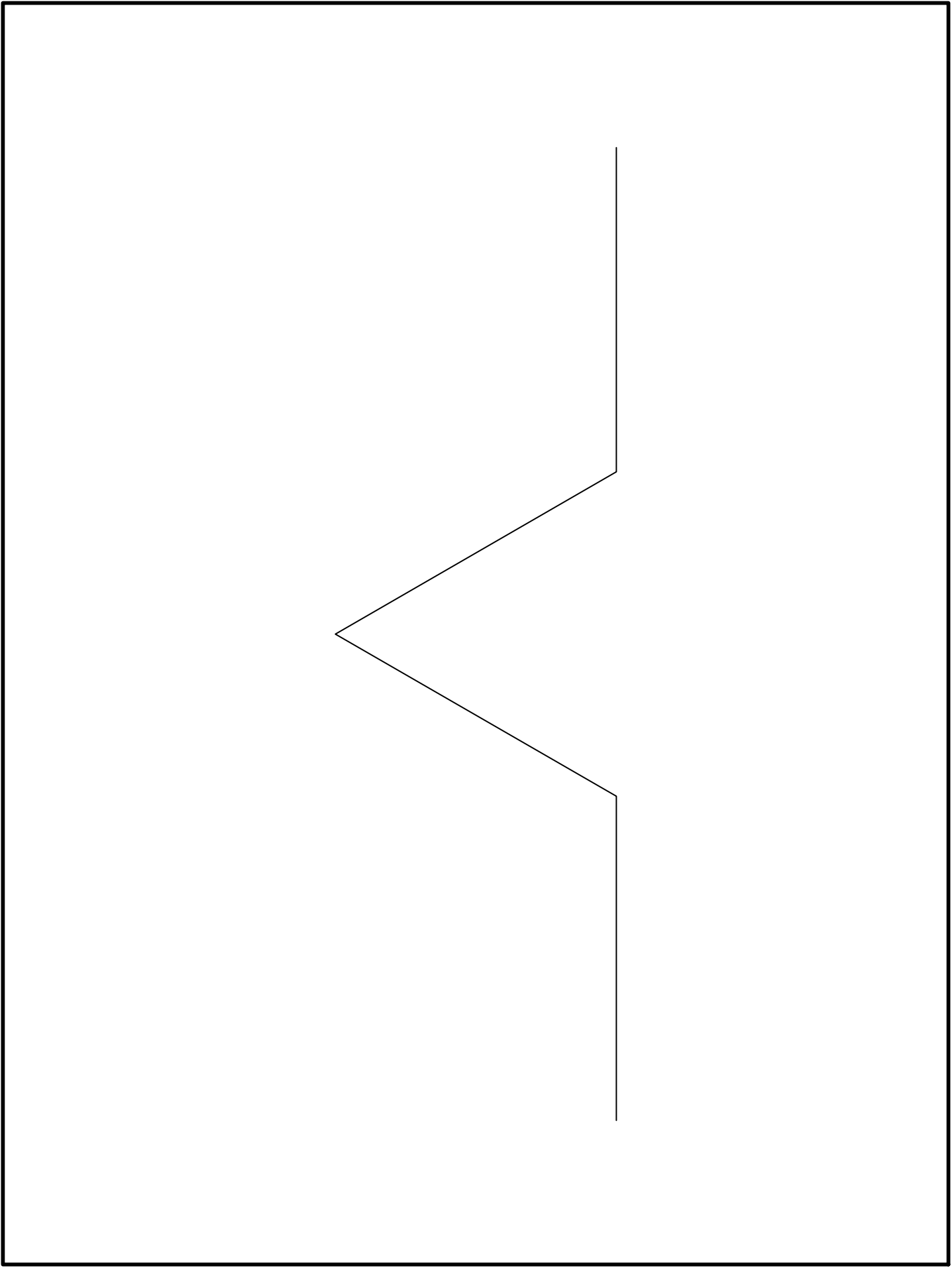
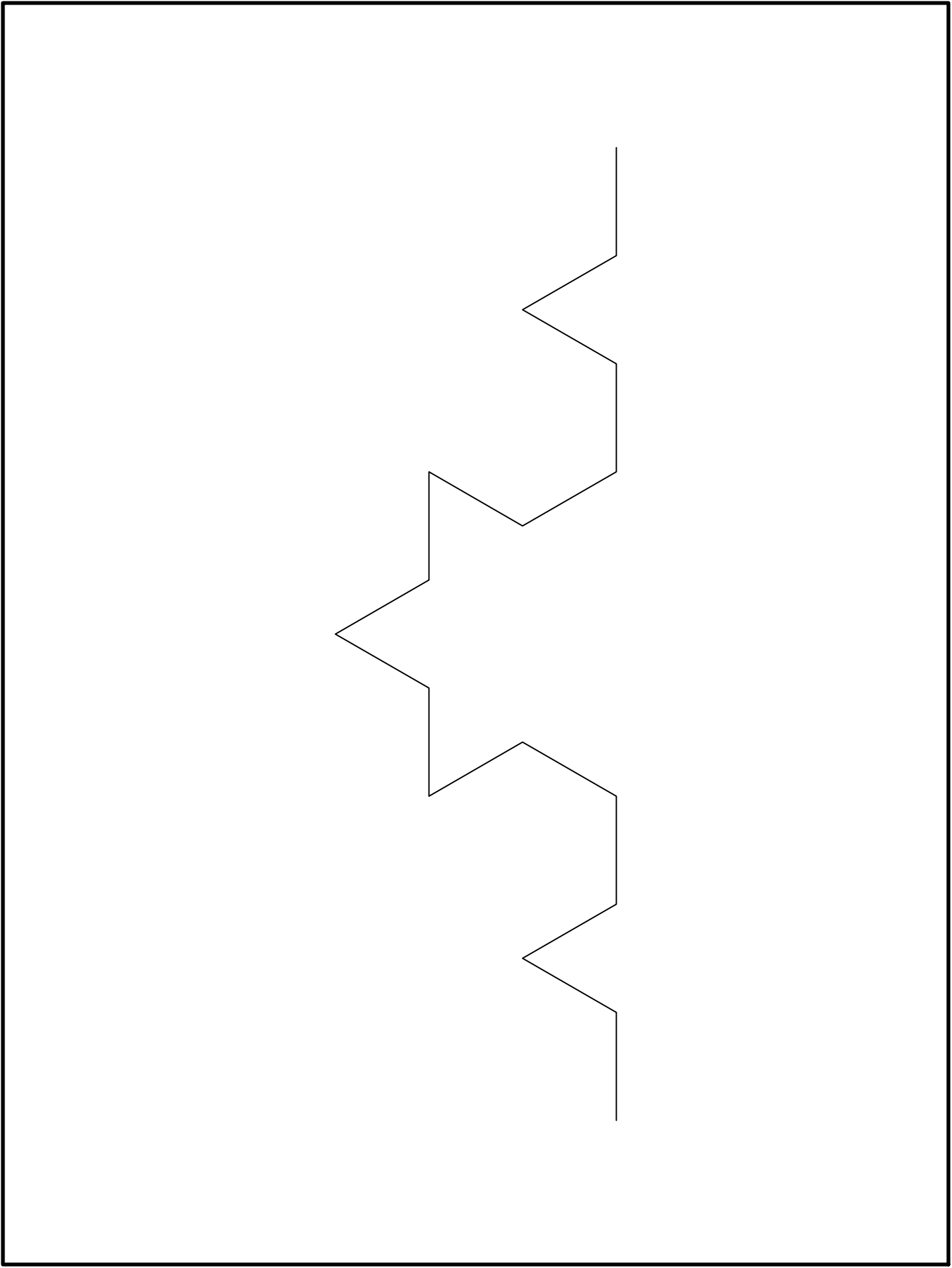
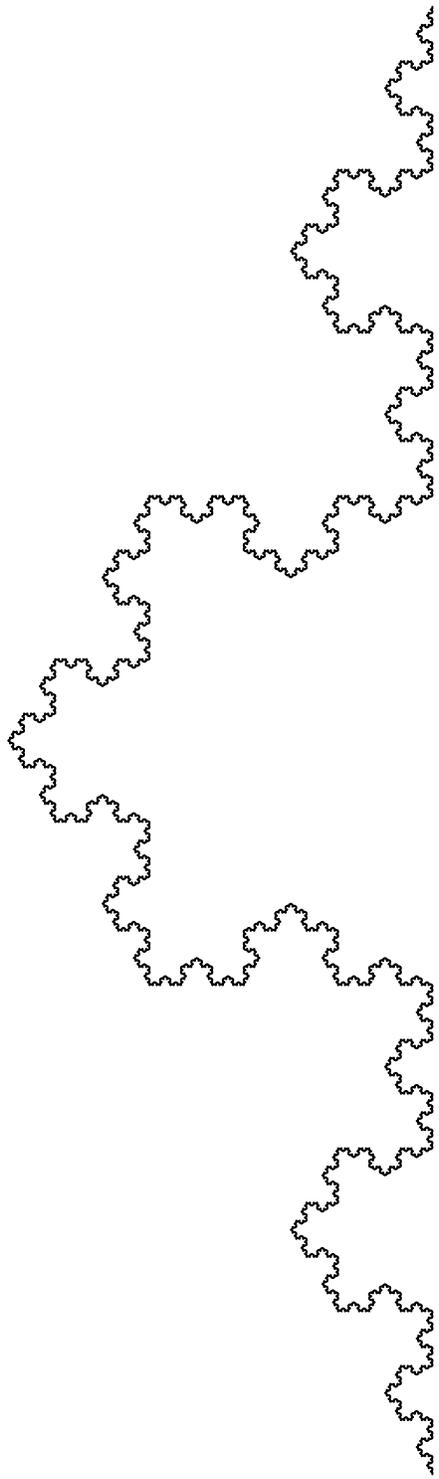
\end{document}